\documentclass[pra,aps,twocolumn,10pt,showpacs,nofootinbib]{revtex4-1}
\usepackage[utf8]{inputenc}
\usepackage[T1]{fontenc}
\usepackage{amsmath}
\usepackage{amsfonts}
\usepackage{amssymb}
\usepackage{graphicx} 
\usepackage{color}
\usepackage{siunitx}
\usepackage{bm}
\usepackage{tikz}
\usepackage[caption=false]{subfig}
\newcommand{\imag}{\mathrm{i}}
\newcommand{\e}{\mathrm{e}}

\newcommand{\PT}{$\mathcal{PT}\,$}

\begin{document}

\title{A Passive $\mathcal{PT}$-Symmetric Floquet-Coupler}
\author{Lucas Teuber}
\author{Florian Morawetz}
\author{Stefan Scheel}
\email{stefan.scheel@uni-rostock.de}
\affiliation{Institut f\"ur Physik, Universit\"at Rostock, 
Albert-Einstein--Stra{\ss}e 23-24, D-18059 Rostock, Germany}

\date{\today}

\begin{abstract}
Based on a Liouville-space formulation of open systems, we present a solution 
to the quantum master equation of two coupled optical waveguides with varying 
loss. The periodic modulation of the Markovian loss of one of them yields a 
passive \PT-symmetric Floquet system that, at resonance, shows a strong 
reduction of the required loss for the \PT symmetry to be broken. We showcase 
this transition for a multi-photon state, and we show how to physically 
implement the modulated loss with reservoir engineering of a set of bath modes.
\end{abstract}


\maketitle

Non-Hermitian systems are an extension to the conventional Hermitian theory 
that provides the basis for our current understanding of quantum physics.
As such, they garnered growing interest in recent years as they promise 
new and exciting ideas and applications. Of special interest are the so-called 
parity-time (\PT) symmetric systems whose non-Hermitian Hamiltonians can still 
have real eigenvalues \cite{BenderBoettcher}.
Depending on the specific system parameters, a phase transition to a regime 
with broken \PT symmetry can be observed where the spectrum becomes complex.
This transition is marked by an exceptional point (EP) where both the 
eigenvalues as well as the corresponding eigenvectors coalesce \cite{Heiss}.

Due to this peculiar behaviour, which results from the complex extension of the 
parameter space, many novel concepts where conceived and implemented. For 
example, the square-root dependence on small deviations for the eigenvalues near 
the EP are thought to lead to increased sensitivity compared to the linear 
dependence of Hermitian degeneracies \cite{Hodaei}. Also, due to the specific 
topology of the non-Hermitian spectrum, a chiral mode-switching can be observed 
when encircling the EP \cite{Doppler}.

Experimental tests of these concepts have already been performed in classical 
setups, e.g. in microwave cavities \cite{Dembowski}, LRC circuits 
\cite{Schindler} or in classical optics \cite{Rueter,Chen,Miri}.
These are mostly active two-mode systems with the loss in one mode being 
balanced by an equal gain in the other. Recently, the first quantum 
experiments have been performed on \PT symmetric systems which showed the 
successful implementation of a \PT directional coupler in integrated waveguides 
\cite{Klauck} and a full quantum-state tomography of a qubit over the EP 
\cite{Murch}. The main difference to classical realisations is that one has to 
implement \PT symmetry passively so as to avoid additional gain noise that 
breaks the \PT symmetry \cite{Scheel}.
However, it was shown that an all-loss passive system can be modelled as an 
active \PT-symmetric system plus an overall loss prefactor \cite{Teuber}
when postselecting on the subspace with highest photon number.

The need for passive \PT systems does have its limits when testing the physics 
at the EP. Experimental implementations with sufficient visibility are hard
to achieve due to the strong losses required to reach the EP and the associated 
low success probability of postselection \cite{Busch}. As there are still many 
questions to be answered, for example, whether in the quantum domain a real 
increase in sensitivity can be achieved, or whether this is off-set by 
quantum noise induced by the self-orthogonality of the coalescing eigenstates 
\cite{Wiersig,Vahala}, a setup is needed that allows to test EP physics with 
reduced losses.

In this Letter, we show that this can be achieved by introducing periodic 
modulation into the system. Based on Floquet theory \cite{Holthaus} one can 
show that the \PT symmetry-breaking threshold is greatly reduced when the 
modulation frequency equals the system's eigenfrequency 
\cite{LeeJoglekar,GundersonJoglekar}.
We calculate the \PT phase diagram of a two-mode waveguide system with 
modulated loss by solving the associated quantum master equation in Liouville 
space utilising a Wei-Norman expansion \cite{WeiNorman}.
A phase transition at a much reduced loss rate amplitude is then illustrated by 
the occupation of multiphoton Fock states. The required modulated loss can
be implemented by a collection of auxiliary waveguides \cite{Meany} that act as
a reservoir simulating the Markovian loss \cite{SzameitLonghi}.


The system under study is comprised of two waveguides with coupling
rate $\kappa$ and a loss rate $\gamma$ for one of them. Both waveguides support 
a single mode described by bosonic operators $\hat{a}_i$ ($i=1,2$). The 
evolution of a quantum state along the propagation direction of the lossy 
waveguides is given by a Lindblad master equation
\begin{equation}
\label{eq:MasterEqHilbert}
\frac{\mathrm{d}}{\mathrm{d}z} \hat{\rho} = - \imag \left[ \hat{H} , \hat{\rho} \right] 
+ \gamma \left( 2 \, \hat{a}_1 \hat{\rho} \hat{a}_1^\dagger - \hat{a}_1^\dagger 
\hat{a}_1 \hat{\rho} - \hat{\rho} \hat{a}_1^\dagger \hat{a}_1 \right),
\end{equation}
with the system Hamiltonian
$\hat{H}=\kappa (\hat{a}_1^\dagger\hat{a}_2+\hat{a}_2^\dagger\hat{a}_1)$
describing a lossless coupler. The modulation of the loss rate $\gamma$ is 
assumed to be slow enough to regard the loss process as being Markovian with a 
dissipator of Lindblad form. We will later discuss this limitation in 
connection with our proposed implementation.

Equation~\eqref{eq:MasterEqHilbert} can be solved in Liouville space 
\cite{Ban}. 
A Liouville space $\mathfrak{L}$ is defined as the Cartesian product 
$\mathfrak{L} = \mathcal{H} \otimes \mathcal{H}'$ of two Hilbert spaces
and amounts to a vectorisation of Hilbert space operators. 
For example, the density operator $\hat{\rho}$ becomes a vector $| \hat{\rho} 
\rangle \rangle$ in Liouville space.
The master equation \eqref{eq:MasterEqHilbert} then reads as
\begin{equation}
\label{eq:LiouvilleSpaceVonNeumann}
\frac{\mathrm{d}}{\mathrm{d}z} | \hat{\rho} \rangle \rangle = \mathcal{L} 
|\hat{\rho} \rangle\rangle,
\end{equation}
where $\mathcal{L}$ is the Liouvillian acting on the original Hilbert 
space operators.
A right-action of the Liouvillian can be defined by introducing the 
superoperators
\begin{gather}
L_k^- | \hat{A} \rangle \rangle 
= \hat{a}_k \hat{A}, \qquad L_k^+ | \hat{A} \rangle \rangle 
= \hat{a}_k^\dagger \hat{A}, \\
R_k^- | \hat{A} \rangle \rangle 
= \hat{A} \hat{a}_k^\dagger, \qquad R_k^+ | \hat{A} \rangle \rangle 
= \hat{A} \hat{a}_k,
\end{gather}
for some Hilbert space operator $\hat{A}$. They inherit commutation relations 
from the bosonic mode operators as $[L_i^-,L_j^+] =[R_i^-,R_j^+] = \delta_{ij}$.
The Liouvillian thus reads
\begin{align}
\label{eq:L}
\mathcal{L} =&- \imag \, \kappa \left( L_1^+ L_{2}^- + L_{2}^+ 
L_1^- - R_{1}^+ R_2^- - R_2^+ R_{1}^- \right)
 \nonumber\\
& + 2 \gamma L_1^- R_1^- -\gamma \left( R_1^+R_1^- + L_1^+L_1^-\right).
\end{align}
Based on this superoperator form, one can employ Lie algebra techniques to 
obtain the evolution superoperator $\mathcal{U}(z)$ that propagates the quantum 
state as
\begin{equation}
\label{eq:evolution}
|\rho(z)\rangle\rangle=\mathcal{U}(z,0)|\rho(0)\rangle\rangle. 
\end{equation}
Here, we employ a Wei-Norman expansion \cite{WeiNorman} of $\mathcal{U}(z)$ 
whose key steps are to first define a Lie algebra 
$\{ X_k \}$ based on the superoperators occuring in Eq.~\eqref{eq:L} and then 
to expand $\mathcal{U}(z)$ as a product of their exponentials, i.e.
\begin{equation}
 \label{eq:WeiNormanExpansion}
 \mathcal{U}(z) = \prod_{k=1}^n \mathcal{U}_k(z) = \prod_{k=1}^n \exp \left[ g_k(z) X_k \right].
\end{equation}
Together with Eqs.~\eqref{eq:LiouvilleSpaceVonNeumann} and \eqref{eq:evolution},
this results in a set of generally nonlinear differential equations 
for the  $g_k(z)$.

Inspecting the Liouvillian \eqref{eq:L}, the Lie algebra is spanned by 
$\{ L^-_i R^-_j, R^+_i R^-_j, L^+_i L^-_j \}$ ($i,j=1,2$),
where the additional operators not present in $\mathcal{L}$ 
are added to close the algebra under commutation. The procedure can now be 
reduced to two separate problems because any Lie algebra can be separated into 
a semisimple and a solvable subalgebra \cite{Gilmore}. In the Wei-Norman 
expansion, this means that the total evolution can be separated into 
$\mathcal{U} =\mathcal{U}_\text{S} \mathcal{U}_\text{R}$.
As the solvable algebra always results in a set of directly integrable 
linear differential equations for the expansion functions, this greatly 
simplifies the overall computation. 
Here, the solvable subalgebra is comprised of
$\{ L_i^- R_j^- \}  \oplus \{\sum_k L_k^+L_k^- , \sum_k  R_k^+R_k^-\}$,
and the semisimple subalgebra is a direct sum of 
two simple algebras
$\{ L_{k}^+L_{k}^- - L_{k+1}^+L_{k+1}^-, L_i^+L_{j\neq i}^- \} 
\oplus  \{ R_{k}^+R_{k}^- - R_{k+1}^+R_{k+1}^-, R_i^+R_{j\neq i}^-\}$.
Due to this separation, we can further decompose
$\mathcal{U}_\text{S}=\mathcal{U}_\text{S$_1$} 
\mathcal{U}_\text{S$_2$}$.
Note that each simple algebra is isomorphic to the special linear algebra 
$\mathfrak{sl}(2,\mathbb{C})$.
The evolution superoperators are thus expanded as
\begin{gather}
\mathcal{U}_\text{S$_1$} = \mathrm{e}^{f_+ L_1^+L_2^-} 
\mathrm{e}^{f_0 \left( L_1^+L_1^- -L_2^+L_2^-\right)} 
\mathrm{e}^{f_- L_2^+L_1^-},\label{eq:US1solution}
\\
\mathcal{U}_\text{S$_2$} = \mathrm{e}^{f_+^* R_1^+R_2^-} 
\mathrm{e}^{f_0^* \left( R_1^+R_1^- - R_2^+R_2^- \right)} 
\mathrm{e}^{f_-^* R_2^+R_1^-},
\label{eq:US2solution}
\\
\mathcal{U}_\text{R} = \mathrm{e}^{a_1 (z) 
\left( L_1^+L_1^- + L_2^+L_2^- \right)} \mathrm{e}^{a_2 (z) 
\left( R_1^+R_1^- + R_2^+R_2^- \right)} \nonumber
\\
\times\mathrm{e}^{a_3 (z) L_1^- R_1^-} \mathrm{e}^{a_4 (z) L_2^- R_2^-}
\mathrm{e}^{a_5 (z) L_2^- R_1^-} \mathrm{e}^{a_6 (z) L_1^- 
R_2^-}.\label{eq:URsolution}
\end{gather}
Inserting the ansatz for the evolution superoperator $\mathcal{U}(z)$ into 
Eqs.~\eqref{eq:LiouvilleSpaceVonNeumann} and \eqref{eq:evolution}
yields the two sets of differential equations for the functions $f_i$ and $a_i$ 
\cite{Korsch,Teuber}.


We now assume the Liouvillian \eqref{eq:LiouvilleSpaceVonNeumann} to be 
periodic, $\mathcal{L}(z)= \mathcal{L}(z+T)$, in order to potentially reduce 
the \PT-breaking threshold. According to Floquet theory, the periodicity 
carries over to the evolution superoperator $\mathcal{U}$ which then obeys 
$\mathcal{U}(z+T) = \mathcal{U}(z) \mathcal{U}(T)$ \cite{Holthaus}.
This means that knowledge of the one-cycle evolution $\mathcal{U}(T)$ 
(the monodromy) allows to construct the evolution for arbitrary $z$. 
The eigenvalues of the monodromy can be written as $\e^{\mu_n T}$ with 
$\mu_n$ being the Floquet exponents whose real parts are the Lyapunov 
exponents that indicate the stability of periodic systems.
Based on the Lyapunov exponents one can decide whether a lossy 
system is \PT-symmetric, or whether that symmetry is broken.
A $z$-independent, \PT-symmetric system is expected to have real 
eigenenergies.
In the \PT-broken phase these eigenvalues becomes complex.
In Liouville space, this behaviour is reversed as the imaginary unit from 
the Schr\"odinger equation has been absorbed in the Liouvillian.
These arguments can be directly transferred to passive periodic systems meaning 
that, if the Lyapunov exponents only show an overall loss of the passive 
system, then \PT symmetry is preserved.
In contrast, when the Lyapunov exponents split from the mean losses, 
\PT-symmetry is broken.

For the passive Floquet \PT coupler we assume the loss in the first waveguide 
to be a periodic function with period $T=2 \pi/\omega$, and to be of the form
\begin{equation}
 \label{eq:gammaFct}
 \gamma (z) = \frac{2 B^2 \exp[-\beta (1-\cos \omega z)]}{\sqrt{1 - B^2 
\exp[-\beta (1-\cos \omega z)]}}.
\end{equation}
Its maximum and minimum values depend on the parameters $B$ and $\beta$,
and we chose the minimum to be $\gamma_\text{min} \approx 0$. 
Its maximum will be denoted by $\bar{\gamma}$.
Inserting the loss rate $\gamma(z)$ and the coupling strength 
$\kappa$ into the differential equations for the functions $f_i$ and $a_i$, we 
numerically compute them up to the period $T$.
Inserted into Eqs.~\eqref{eq:US1solution}--\eqref{eq:URsolution} gives the
evolution superoperator $\mathcal{U}(T)$, from which the Lyapunov exponents are
obtained by diagonalisation.

\begin{figure}
\includegraphics[width=0.4\textwidth]{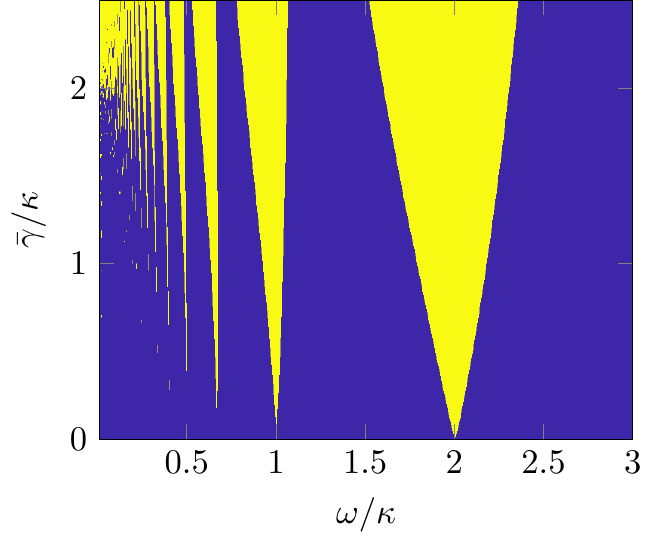}
\caption{\PT phase diagram of the passive coupler with modulated loss 
$\gamma (z)$ from Eq.~(\ref{eq:gammaFct}). For Lyapunov exponents that split 
from the mean loss values, the system is in the \PT-broken phase (yellow).}
\label{fig:PTphaseDiagram}
\end{figure}

As an example, we consider a single photon propagating through the two-mode 
waveguide system. In Fig.~\ref{fig:PTphaseDiagram}, the resulting \PT phase 
diagram is shown as a function of the modulation frequency $\omega$ and the 
maximum loss $\bar{\gamma}$, normalized with respect to the coupling constant 
$\kappa$ between the waveguides. The \PT-broken phase is shaded in yellow.
The diagram shows a clear reduction of the \PT-breaking threshold at the 
resonance frequency $\omega = 2 \kappa$ of the lossless coupler, with
additional regions of reduced thresholds for lower modulation frequencies.
A similar behaviour was also observed in a different context 
\cite{LeeJoglekar}, thus pointing at a universal behaviour.
Preparing the passive system with a loss modulation frequency equal to the 
resonance frequency might therefore enable one to efficiently probe the
transition between \PT-symmetric and broken phases.

Note that the \PT phase diagram as calculated in Liouville 
space is identical to the Hilbert space phase diagram calculated from an 
effective non-Hermitian Hamiltonian of an active two-mode \PT system.
If this were not true, the passive system would not be viable to simulate 
the active system. That this is indeed the case can be deduced from the 
decomposition of the algebra and the subsequent product form of the evolution 
superoperators. The superoperators $L_i^- R_j^-$ that are responsible for 
removing photons, as well as the sum of superoperators responsible for the mean 
loss, are clearly separated from the $\mathfrak{sl}(2,\mathbb{C})$ algebras 
that describe the underyling active \PT coupler.
When postselecting on the outcome where no photon is lost in transmission, the 
contributions from $L_i^- R_j^-$ can be dismissed, and the only remaining part 
is an evolution governed by the effective non-Hermitian Hamiltonian of the 
active \PT system plus an overall mean loss.
However, this mean loss is now greatly reduced at the 
resonance frequency $\omega=2\kappa$ compared to the unmodulated case where the 
threshold is $\bar{\gamma} = 2 \kappa$. Additionally, because we already solved 
for the monodromy, we are also able to calculate the full quantum 
evolution over all subspaces without the need for postselection.

\begin{figure*}[t]
\includegraphics[width=8.5cm]{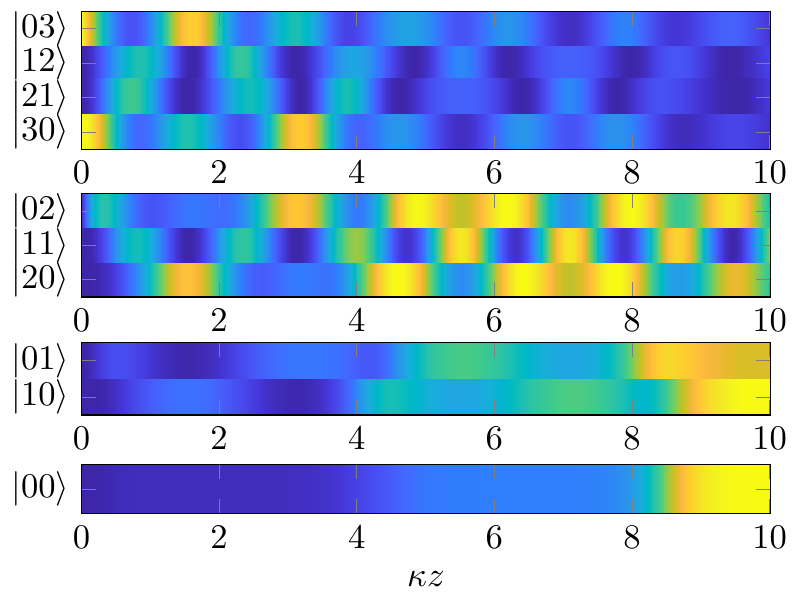} \hfill
\includegraphics[width=8.5cm]{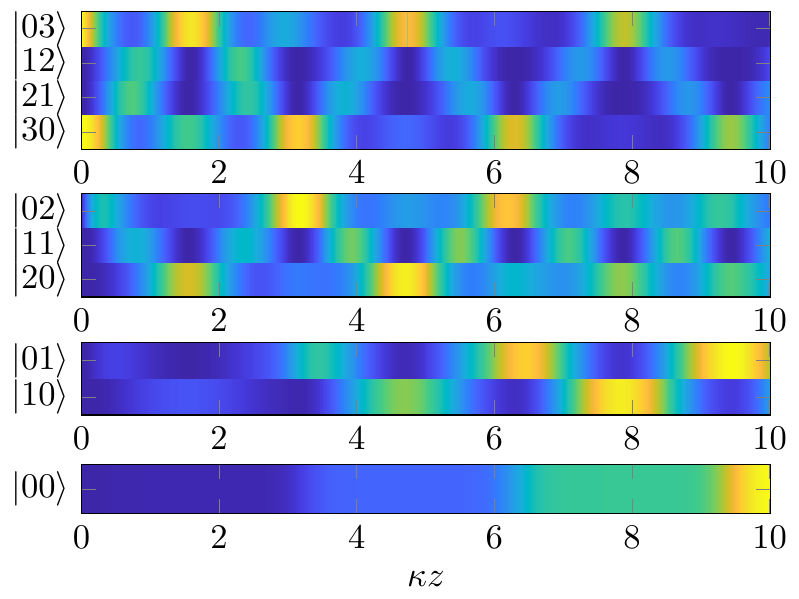}
\caption{Occupations $P(n, h; z)$ for states $| n-h, h \rangle$ with input 
state $|\psi\rangle = (|0, 3 \rangle + |3, 0\rangle)/\sqrt{2}$ over all 
subspaces with photon numbers $n \in \{ 0, 1, 2, 3 \}$. (Left) 
\PT-symmetric phase ($\bar{\gamma}=0.25 \kappa$, $\omega=1.5\kappa$). (Right) 
\PT-broken phase ($\bar{\gamma}=0.25 \kappa$, $\omega=2\kappa$).}
\label{fig:compare}
\end{figure*}

We highlight one example of how the \PT phase transition manifests itself by 
calculating the occupation $P(n,h,z)$ of states $|n-h,h\rangle$ 
\begin{equation}
 P (n, h; z) = \langle n-h,h | \hat{\rho}(z) | n-h, h \rangle,
\end{equation}
over different subspaces with photon numbers $n$.
Starting with the input state $|\psi\rangle=(|0,3\rangle+|3,0\rangle)/\sqrt{2}$,
we show in Fig.~\ref{fig:compare} the evolution of all photon-number 
subspaces using a coupling constant $\kappa=1$.
In the left figure, the system has a loss amplitude 
$\bar{\gamma}=0.25 \kappa$ and a modulation frequency $\omega=1.5\kappa$ 
associated with the \PT-symmetric phase (see Fig.~\ref{fig:PTphaseDiagram}), 
whereas in the right figure, the modulation frequency is set to 
$\omega=2\kappa$ associated with the \PT-broken phase.
This is reflected in the general behaviour of the occupations 
$P(n,h,z)$. In the left panel of Fig.~\ref{fig:compare}, there are two 
oscillating strands that are equally damped by an overall loss. In contrast, the 
panel on the right shows one strongly damped strand and one with significantly 
lower loss. This behaviour is repeated across all subspaces except the 
continuously filled vacuum subspace (lowest subpanels). Note also that all 
subspaces with fewer than 3 photons are only 
transiently occupied.

The qualitative difference of the two evolutions is a clear sign of a \PT 
symmetry breaking where the system transitions from a coherent evolution (plus 
overall loss in the passive scheme) to an evolution that splits into 
exponentially decaying and growing modes. The physical explanation for this 
behaviour is that the damping of the $z$-dependent Floquet modes depend on 
whether or not they are concentrated in states $|n-h,h\rangle$ with more 
photons in the lossy waveguide when $\gamma(z)$ is large.
This is the Floquet analogue of the usual signature of broken \PT
symmetry of one mode being amplified and the other one being suppressed.

Recall that this \PT-symmetry breaking is only initiated by a change of the 
modulation frequency $\omega$, and that the loss amplitude $\bar{\gamma}$ is 
held at a low and constant value. In the static case, one instead has to change 
the loss rate to higher values that lead to significantly reduced visibilities 
in the measurements. The passive Floquet \PT coupler is therefore a possible 
way to probe the \PT phase transition without the obstacle of the overall 
loss. This is especially interesting as the required loss rate might even be 
further reduced as seen from the phase diagram Fig.~\ref{fig:PTphaseDiagram}.
However, as the range of frequencies, for which \PT-symmetry is broken, 
becomes progressively narrower with decreasing values of $\bar{\gamma}$, an 
experimental implementation becomes more challenging.

\begin{figure}[h]
\includegraphics[width=0.45\textwidth]{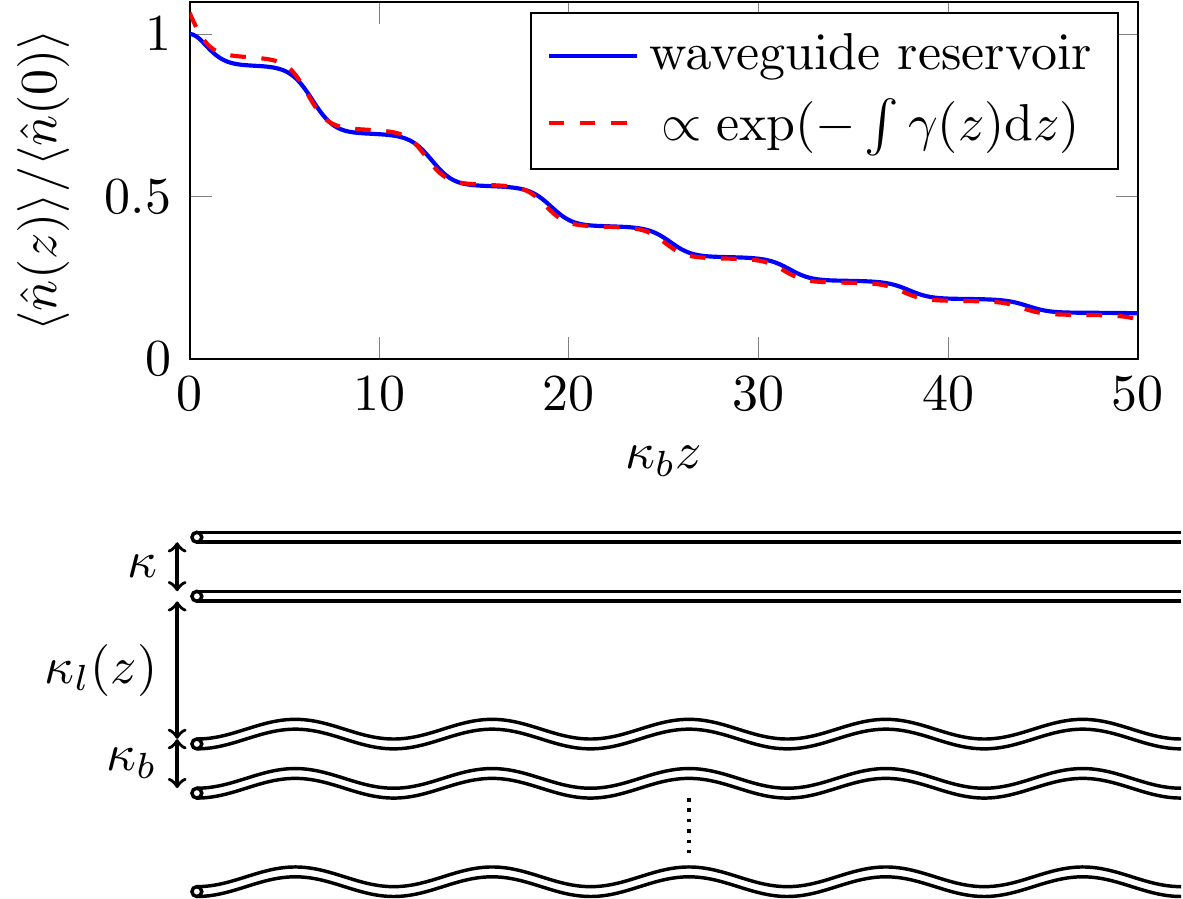}
\caption{(Top) Decay of the mean photon number in the system waveguide 
by coupling to a modulated reservoir of waveguides (blue solid line) and 
comparison to analytical form of decay (red dashed line).
(Bottom) Sketch of the modulated reservoir coupled to system waveguides.}
\label{fig:mod.decay}
\end{figure}

Finally, we present a proposal on how to implement such a lossy coupler using 
auxiliary waveguides. The general principle is depicted in the lower part of 
Fig.~\ref{fig:mod.decay}. The top pair of waveguides, 
together with their mutual coupling $\kappa$, constitute the system under 
investigation. The lower waveguide of the pair is additionally coupled to a 
homogeneous array of $N$ auxiliary waveguides (the reservoir) with the coupling 
$\kappa_l$, while the coupling inside the reservoir is denoted by $\kappa_b$.
In order to concentrate on the loss implementation, we briefly consider only 
one active system waveguide ($\kappa = 0$).
In the weak-coupling regime where $\kappa_l\ll\kappa_b$, the population in the 
system waveguide approximately shows an exponential decay with rate
\begin{equation}
\label{eq:decay}
\gamma=\frac{2 \kappa^2_l}{\sqrt{\kappa_b^2-\kappa^2_l}}
\end{equation}
after some short initial parabolic decay \cite{Longhi06}. For $N\to\infty$, 
the lost population does not return to the system waveguide, and hence
constitutes a Markovian loss. For finite $N$, the exponential decay is
only a good approximation up to some recurrence time due to reflections at the 
end of the array, which scales linearly with the array size. However, a 
sufficient number of auxiliary waveguides is easily obtainable in experiments 
\cite{Dreisow08}.

A modulated loss can then be implemented by modulating the coupling $\kappa_l$ 
which, in the evanscent coupling of the intregated photonic waveguides has the 
general form $\kappa_l(z)=A\exp[-\alpha d(z)]$, where $d(z)$ is the distance 
between waveguides and $A$ and $\alpha$ 
are appropriate scaling factors. With a modulation function
$d(z)\propto\frac{1}{2}(1-\cos{(\omega z)})$, Eq.~\eqref{eq:decay}
yields the loss rate in Eq.~\eqref{eq:gammaFct}. Note that the 
modulation has to be sufficiently slow for the resulting decay to follow an 
exponential law, i.e. that it can be described by a rate that yields the correct 
form of the dissipator of the quantum master equation 
\eqref{eq:MasterEqHilbert}.

In order to check the validity of our assumptions, we compared the 
numerical evaluation of the $N+1$ waveguide model with the behaviour of a lossy 
waveguide with a modulated decay rate ($\propto \exp{(-\int 
\gamma(z)\mathrm{d}z)}$ with $\gamma(z)$ given by Eq.~\eqref{eq:gammaFct}.
The result is shown in the upper panel in Fig.~\ref{fig:mod.decay} for 
$\bar{\gamma}=0.125 \kappa_b$ and $\omega=\kappa_b$.
Setting $\kappa = 0.5\kappa_b$ this corresponds to the example of the 
\PT-broken phase with $\bar{\gamma}=0.25 \kappa$ and $\omega=2\kappa$ 
(right panel in Fig.~\ref{fig:compare}).
Note that the initial parabolic decay in the analytical approximation (dashed 
line) was accounted for by appropriate normalization \cite{Longhi06}.
The numerical result (solid line) for the $N+1$ waveguide system matches 
the exponential decay with modulated frequency [Eq.~\eqref{eq:decay}] very well.
After re-introducing the system coupling $\kappa > 0$, the loss still follows 
the exponential decay very closely, thus enabling the simulation of the Floquet 
\PT coupler.


In conclusion, we presented a method to probe the \PT-breaking transition in a 
passive Floquet \PT coupler with a modulated loss rate $\gamma (z)$. The \PT 
phase diagram was calculated for a functional form of the loss rate suitable 
for the implementation using evanscently coupled photonic waveguides.
We showed that a phase transition occurs at considerably lower loss rates 
compared to the static case, which provides a feasible route to study 
\PT-symmetry breaking in quantum optical systems, in which modulated losses 
can be tailored by reservoir engineering.

This work was supported by the Deutsche Forschungsgemeinschaft (DFG) through 
grant SCHE 612/6-1.

\end{document}